\documentclass[english,twocolumn,prl]{revtex4}
\usepackage[T1]{fontenc}
\usepackage[latin9]{inputenc}
\usepackage{bm}
\usepackage{amsmath}
\usepackage{amssymb}
\usepackage{lipsum}
\usepackage{graphicx}
\usepackage{bm}
\usepackage{ dsfont }
\usepackage{makeidx}
\usepackage{epstopdf} 
\usepackage[retainorgcmds]{IEEEtrantools}
\usepackage{hyperref} 
\usepackage{braket}

\def\to{\rightarrow}

\newcommand{\beq}{\begin{equation}} 
\newcommand{\eeq}{\end{equation}} 
\newcommand{\beqa}{\begin{eqnarray}} 
\newcommand{\eeqa}{\end{eqnarray}}


 
\usepackage{babel}
\begin{document}
\bibliographystyle{naturemag}

\title{Anisotropic fixed points in Dirac and Weyl semimetals. }
\author{\'Oscar Pozo$^1$, Yago Ferreiros$^2$,  and  Mar\'ia A. H. Vozmediano$^1$ }
\affiliation{ $^1$ Instituto de Ciencia de Materiales de Madrid, and CSIC, Cantoblanco, 28049 Madrid, Spain}
\affiliation{$^2$ Department of Physics, KTH Royal Institute of Technology, SE-106 91 Stockholm, Sweden  }

\begin{abstract}
The effective low energy description of interacting Dirac and Weyl semimetals is that of massless Quantum Electrodynamics with several Lorentz breaking material parameters. 
We perform a renormalization group analysis of Coulomb interaction in anisotropic Dirac and Weyl semimetals and show that the anisotropy persists in the material systems at the infrared fixed point. In addition, a tilt of the fermion cones breaking inversion symmetry induces a magneto--electric term in the electrodynamics of the material whose magnitude runs to match that of the electronic tilt at the fixed point. 
\end{abstract}
\maketitle

\section{Introduction}
\label{sec_intro}
The experimental realization of Dirac and Weyl semimetals (WSM) in three dimensions (3D)\cite{BGetal14,LJetal14,Xu15,Lvetal15,Xuetal15,Yanetal17} has opened a new avenue in condensed matter physics in the post--graphene era. The  attraction of these materials comes in part from the shared properties with their high energy counterparts and the fruitful in-breeding that their studies bring to both communities. 

The effective low energy model of interacting WSMs is very similar  to massless (zero fermion mass) Quantum Electrodynamics (QED). The breakdown of Lorentz symmetry induced by various material parameters (the separation of the Weyl nodes, the departure of the  Fermi velocity from the speed of light $c$   and, eventually, the tilt of the cones) does not alter significantly the renormalization properties of the system. In particular, its most relevant feature, the  infrared stable fixed point, survives in these Lorentz breaking models. 
Lorentz invariance violating terms (LIV) in pure QED have been extensively explored in  the context of Quantum Field Theory \cite{CFJ90,CK98,KM09,JK99} where astrophysical observations put severe bounds to their presence \cite{KR11}.  A renormalization group (RG) analysis of all the possible LIV breaking parameters was performed in \cite{KLP02} with the finding that full Lorentz invariance (rotations and boosts) is always restored at the infrared critical point. 

In the condensed matter context the main question is the stability and ultimate fate of the system in the infrared limit. Systems with a regular Fermi velocity were analyzed in \cite{P93} and shown to give rise to standard Fermi liquid. Singular Fermi surfaces in 2D were explored in the early publications \cite{GGV94,GGV99} associated to graphene. 
There it was shown that the Fermi velocity grows monotonically till it reaches the speed of light, a result that has been later obtained by many techniques (see a recent account with a fair list of references in \cite{TK18}) and was confirmed experimentally in \cite{EGetal11,MEetal12}. This  result is very robust although the infrared fixed point with $v_F=c$ is experimentally unreachable \cite{JGV10,V11}. A very complete  analysis of the 2D model including short range interactions has appeared in the recent publication \cite{TLetal18}.

RG phases of the 3D massless QED problem in a condensed matter framework were explored in the early works \cite{AB71,IN12} prior to the experimental realization of Dirac and Weyl semimetals (see also \cite{RJH16}). The non--relativistic limit with a static Coulomb potential was later addressed in \cite{RL13,Sarma15,DFG17}. A crucial difference with the 2D case is that the polarization diagram is finite in the 2D case and does not induce a renormalization of the permeability or susceptibility. The speed of light and the electric charge are not renormalized and the RG running of the effective coupling constant  comes solely from the Fermi velocity renormalization. The same happens in 3D with the static Coulomb propagator. When taking into account the full retarded photon propagator in 3D  the divergent polarization diagram renormalizes the velocity of light through the electric permeability and the magnetic susceptibility. The full relativistic isotropic case was analyzed in  \cite{AB71,IN12}; it was found that both the fermion and photon velocities run to a common, isotropic and non--universal value at the infrared fixed point. This restores Lorentz symmetry, in particular it allows to define a single Lorentz factor. Rotational invariance was not questioned in these works. The material  realizations show a Fermi velocity anisotropy and tilt, both affecting the $SO(3)$ rotational symmetry, a part of the Lorentz group. 

In this work we analyze the renormalization of the various parameters of the WSM interacting model using the full Coulomb interaction mediated by relativistic photons. In particular we analyze the case of an initially anisotropic dispersion relation and a tilt. Similarly to what happens in the isotropic case discussed in \cite{IN12,RJH16}, we find that anisotropic fermion and photon velocities also run to a common, non--universal value at the fixed point. Hence Lorentz boosts can be defined at each particular direction but rotational symmetry is not recovered at the infrared fixed point. 
We also find that a tilt $t$ in the matter sector that breaks inversion symmetry ($\mathcal{I}$) induces a magneto--electric coupling in the WSM electrodynamics whose value runs to a common, non--vanishing value with $t$ in the infrared fixed point. Contrary what happens in the static limit \cite{DFG17}, in this fully relativistic analysis, the tilt does not vanish in the infrared fixed point.

In contrast to the high energy context, LIVs terms of WSMs are not restricted to very small values and their experimental accessibility selects a preferred frame and allows for the anisotropic fixed point described in this work.
In particular, the Fermi velocity and tilt of the interacting fermionic system can be directly observed in angle resolved photoemission (ARPES) experiments \cite{CSetal16,DWetal16,JLetal17,LWetal17}, providing physical initial values for the RG flow ending in the anisotropic fixed points. Standard optical probes can also reveal birefringence associated to the time--reversal ($\mathcal{T}$) breaking tilt term.

We do not address the role of short range interactions, disorder, or non--perturbative effects leading to spontaneous symmetry breaking; some of these issues have been explored in   \cite{WKA14,JG15,JG17,RGJ17,SF17}. The vector $b_{\mu}$ separating the Weyl cones in $\mathcal{T}$ broken WSMs does not alter our results. Finally, the term ``anisotropic WSMs" often refers to systems where the electronic dispersion around a Weyl point is linear in some directions and quadratic in others. The RG analysis is different in these cases \cite{LWL18}.

\section{The model}
\label{sec_model} 
Weyl fermions in WSMs can be described by a LIV extension of massless QED.
Depending on the tilt of their energy cone, WSMs are classified as  Type I and Type II \cite{SGetal15}. The Fermi surface of Type I is a point, while Type II WSMs do have an extended Fermi surface, the electron--electron interaction is screened and the scaling analysis underlying the RG approach is substantially different (see \cite{P93}) from the one of the Type I case. Hence, we focus only on type I WSMs. A dispersion relation of the form (without loss of generality we choose the tilt in the $Z$ direction) 
\begin{equation}
   E=\pm\sqrt{\sum_{i=1}^3 v_{i}^{2}p_{i}^{2}}+tp_{3}  \  ,
\label{eq_tilt}   
\end{equation}
is obtained from the tree level Lagrangian
\begin{equation}
   \mathcal{L}_{\text{F}}=i\bar{\psi}\left( \gamma^{0}\left( \partial_{0}-t\partial_{3}\right) +\sum_{i=1}^3\gamma^{i}v_{i}\partial_{i} \right) \psi  \  ,
   \label{LF}
\end{equation}
where $\psi$ is the fermionic field, $\gamma^{\mu}$ are the contravariant gamma matrices, $t$ is the tilt velocity that breaks both $\mathcal{I}$ and $\mathcal{T}$ and $v_{i}$ are the components of the Fermi velocity. In our convention the metric is $\eta_{\mu\nu}=\text{diag}\left( 1,-1,-1,-1 \right)$. All the velocities, including the tilt parameter, will be given in units of the speed of light in vacuum, $c_{0}=1$. The condition $|t|<|v_{3}|$ ensuring type I will be kept throughout the work. 

The electromagnetic interaction  is obtained by replacing the ordinary derivative of \eqref{LF} by a covariant derivative
\begin{equation}
   \mathcal{L}_{\text{int}}=-e\bar{\psi}\left( \gamma^{0}\left(A_{0}-tA_{3} \right) +\sum_{i=1}^3\gamma^{i}v_{i}A_{i} \right) \psi  \  ,
   \label{Lint}
\end{equation}
where $e$ is the electric charge and $A_{\mu}$ is the photon field. Finally, we need to construct an appropriate photon propagation. The standard term in QED (in the Lorenz gauge $\partial_{\mu}A^{\mu}=0$ \footnote{This gauge condition is due to L. V. Lorenz, not to be confused with H. Lorentz. We thank an anonymous referee for pointing this to us.})
\begin{equation}
   \mathcal{L}_{\text{ph QED}}=-\dfrac{1}{4}F_{\mu\nu}F^{\mu\nu}-\dfrac{1}{2\xi}\left( \partial_{\mu}A^{\mu} \right)^{2}  \  ,
\end{equation}
where $F_{\mu\nu}=\partial_{\mu}A_{\nu}-\partial_{\nu}A_{\mu}$ is the electromagnetic tensor, is 
too much constrained by Lorentz invariance and does not allow to renormalize the vacuum polarization divergencies arising in the anisotropic WSM. We need a term that reflects the anisotropy of the media, so we introduce a polarization tensor in which the permittivity $\epsilon$ and permeability $\mu$ (both of them will be given in units of the permittivity $\epsilon_{0}$ and permeability $\mu_{0}$ of the vacuum) depend on the direction in which they are measured.
\begin{equation}
   \mathcal{L}_{\text{ph}}=\dfrac{1}{2}\sum_{i=1}^3\left( \epsilon_{i}E_{i}^{2}-\dfrac{1}{\mu_{i}}B_{i}^{2} \right) -\dfrac{1}{2\xi}\left( \partial_{\mu}A^{\mu} \right)^{2}  \  ,
   \label{eq_Lph}
\end{equation}
where $\vec{E}$ and $\vec{B}$ are the electric and magnetic fields defined in terms of the photon field in the standard way:
$	E_{i}=\partial_{0}A_{i}-\partial_{i}A_{0}    , \;  B_{i}=-\epsilon_{ijk} \partial^{j}A^{k}.$
Finally, we will see that a fermion tilt generates additional polarization diagrams that can not be absorbed in the parameters in eq. \eqref{eq_Lph}. If the tilt is chosen to be in the $Z$ direction,  the photon propagator needs a structure analog to the fermion tilt which consists of replacing
\beq
	E_{1}\to E_1 +\omega_1 B_2  \  , \qquad E_2\to E_{2}-\omega_2 B_1  \  ,
	\label{eq_tiltPh}
\eeq
in eq. \eqref{eq_Lph} so that the permeability in the plane perpendicular to the electronic tilt is modified and a linear magneto--electric term is generated. 
\begin{IEEEeqnarray}{rCl}
\mathcal{L}_{\text{ph}} & = & \dfrac{1}{2}\sum_{i=1}^3\left( \epsilon_{i}E_{i}^{2}\right)-\dfrac{1}{\mu_{1}}\left( 1-\dfrac{\omega_{1}^{2}}{c_{1}^{2}} \right) B_{1}^{2} \nonumber \\
& - & 
\dfrac{1}{\mu_{2}}\left( 1-\dfrac{\omega_{2}^{2}}{c_{2}^{2}} \right) B_{2}^{2}-
\dfrac{1}{\mu_{3}}B_{3}^{2}  \nonumber \\
& + & \epsilon_{1}\omega_{1}E_{1}B_{2}-\epsilon_{2}\omega_{2}E_{2}B_{1} -\dfrac{1}{2\xi}\left( \partial_{\mu}A^{\mu} \right)^{2}  \  ,
\label{eq_Lph2}
\end{IEEEeqnarray}
where we have defined  $c_{i}=c_{0}/\sqrt{\epsilon_{i}\mu_{i}}$.
\section{Renormalization of the model}
\label{sec_app} 
\begin{figure}
	\centering
	\includegraphics[scale=0.15]{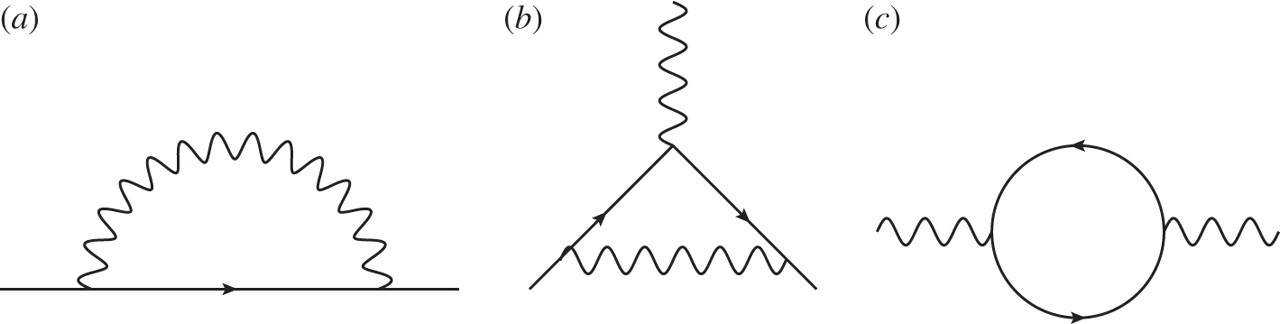}
	\caption{1-loop divergent diagrams of QED. a) Electron self-energy, b) Vertex, c) Vacuum polarization.}
	\label{fig_1loopDD}
\end{figure}
In QED there are three primitively divergent diagrams depicted in Fig. \ref{fig_1loopDD}. For anisotropic WSMs the same three diagrams give rise to various independent divergences due to anisotropy. Potentially divergent one--loop diagrams with with three and four external photon legs arising from LIV QED have been demonstrated to vanish in ref. \onlinecite{KLP02}.

Feynman diagrams of the model are constructed with the following propagators and vertex:
\begin{IEEEeqnarray}{rCl}
	S_{\text{F}}(k) & = & \dfrac{i}{\gamma^{0}\left( k_{0}-tk_{3} \right) +
		\sum_{i=1}^3\gamma^{i}v_{i}k_{i} } \;  ,  \\
	G_{\mu\nu} & = & iM_{\mu\nu}^{-1}  \  ,  \\
	V^{\mu} & = & -iel^{\mu}_{\nu}\gamma^{\nu}  \;  ,  \\
	l^{\mu}_{\nu} & = & \left( \begin{matrix} 1 & 0 & 0 & 0 \\ 0 & v_{1} & 0 & 0 \\ 0 & 0 & v_{2} & 0 \\ -t & 0 & 0 & v_{3}  \end{matrix} \right)  \  ,
	\label{eq_propagators}
\end{IEEEeqnarray}
and the matrix $M_{\mu\nu}$ is the one that appears when  eq. \eqref{eq_Lph2} is rewritten as
\begin{equation}
   \mathcal{L}_{\text{ph}}=\dfrac{1}{2}A^{\mu}M_{\mu\nu}A^{\nu}  \  .
\end{equation}
Symbolically, the electron self-energy, vacuum polarization and vertex diagrams are given by
\begin{IEEEeqnarray}{rCl}
	\Sigma(p)  &=&  \int\dfrac{d^{4}k}{(2\pi)^{4}} \ V^{\mu}S_{\text{F}}(p-k)V^{\nu}G_{\mu\nu}(k)  \  ,  \label{ESE} \\
	\Pi^{\mu\nu}(q)  &=&  -\int\dfrac{d^{4}k}{(2\pi)^{4}} \ \text{Tr}\left[ V^{\mu}S_{\text{F}}(k)V^{\nu}S_{\text{F}}(k-q) \right]  , \label{VP} \\
	\Gamma^{\mu}(0,0) &=& \int\dfrac{d^{4}k}{(2\pi)^{4}} \ V^{a}S_{\text{F}}(k)V^{\mu}S_{\text{F}}(k)V^{b}G_{ab}(k)  \  .  \label{Vertex}
\end{IEEEeqnarray}
A set of 13 counterterms are introduced in the parameters of the lagrangian  to cancel the divergencies of these diagrams. Due to gauge invariance, not all are independent. We choose a renormalization of the polarization tensor such that neither the photon field $A_{\mu}$ nor the electric charge $e$  renormalize. The parameters $\epsilon_i, \mu_i$ determine the renormalization of the speed of light in the different directions. 

\section{Beta functions and results}
\label{sec_beta}
The beta functions of the parameters of the model are defined by
$\beta_x \equiv dx/d(\log\Lambda)$,
where $\Lambda$ is the energy scale introduced in the renormalization procedure. We have used a dimensional regularization scheme to define the counterterms  described in Appendix \ref{sec_ApRen} which
leads to the following beta functions
\begin{widetext}
\begin{IEEEeqnarray}{rCCCCCCCl}
	\beta_{t}  =  \alpha_{3}\epsilon_{3} \left( tF_{0}^{0}+F_{3}^{0} \right)   , \qquad\quad
	&\beta_{v_{i}}= \alpha_{i}\epsilon_{i}\left( v_{i}F_{0}^{0}-F_{i}^{i} \right)  ,  \quad\quad i=1,2,3 
	\label{eq_betat} \nonumber\\    
	\qquad 
	\beta_{\omega_{1}} = \dfrac{2\alpha_{1}}{3}\dfrac{v_{1}}{v_{2}v_{3}}\left( \omega_{1}-t \right)  \  , \quad 
&\beta_{\omega_{2}}=\dfrac{2\alpha_{2}}{3}\dfrac{v_{2}}{v_{1}v_{3}}\left( \omega_{2}-t \right)  \  , \label{eq_betaomega}\\
	\beta_{\epsilon_{1}}  =  -\dfrac{2\alpha_{1}}{3}\epsilon_{1}\dfrac{v_{1}}{v_{2}v_{3}} \ ,\qquad
&	\beta_{\epsilon_{2}}  = -\dfrac{2\alpha_{2}}{3}\epsilon_{2}\dfrac{v_{2}}{v_{1}v_{3}}  \  , \qquad  
&	\beta_{\epsilon_{3}}  =  -\dfrac{2\alpha_{3}}{3}\epsilon_{3}\dfrac{v_{3}}{v_{1}v_{2}} \  , \quad  \nonumber\\
\beta_{\mu_{1}}  =  \dfrac{2\alpha_{1}}{3}\epsilon_{1}\mu_{1}^{2}\dfrac{v_{2}\left( v_{3}^{2}-\left( t-\omega_{2} \right)^{2} \right) }{v_{1}v_{3}} ,	\qquad 
  &  \beta_{\mu_{2}}  =  \dfrac{2\alpha_{2}}{3}\epsilon_{2}\mu_{2}^{2}\dfrac{v_{1}\left( v_{3}^{2}-\left( t-\omega_{1} \right)^{2} \right) }{v_{2}v_{3}}  ,	\qquad  
&	\beta_{\mu_{3}}  =  \dfrac{2\alpha_{3}}{3}\epsilon_{3}\mu_{3}^{2}\dfrac{v_{1}v_{2}}{v_{3}}  \  ,
	\label{eq_betac}
\end{IEEEeqnarray}	
%
\end{widetext}

In what follows we will analyze the RG flows of the most significant examples. The dimensionless coupling constants of our theory are given by $\alpha_{i}\equiv\alpha/\epsilon_{i}$ for $i=1,2,3$.
They  go to zero in the infrared limit  
in all the cases analyzed through the article.

\subsection{Isotropic case}
\begin{figure*}
	\centering
	\includegraphics[width=0.60\columnwidth]{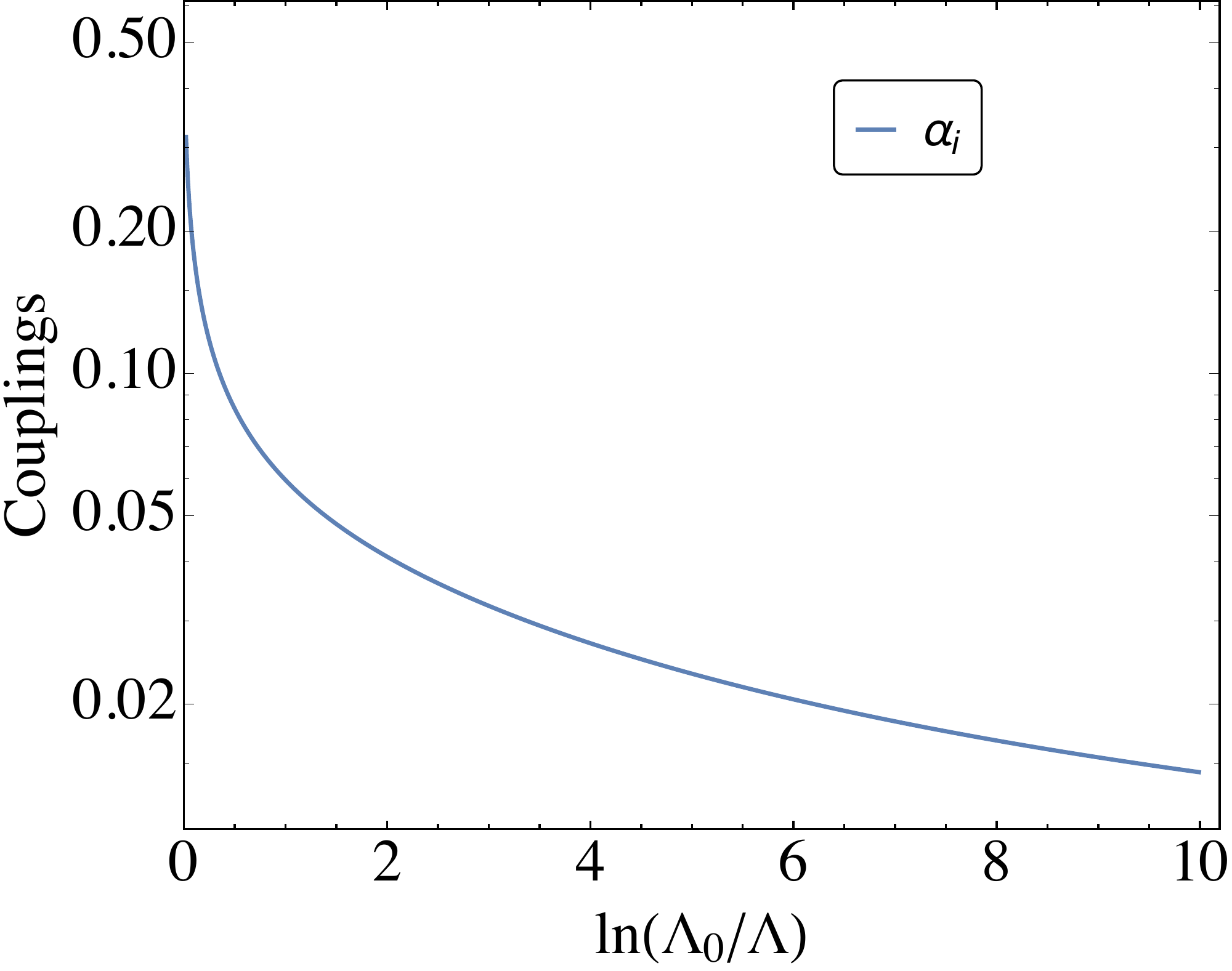}(a)
	\includegraphics[width=0.60\columnwidth]{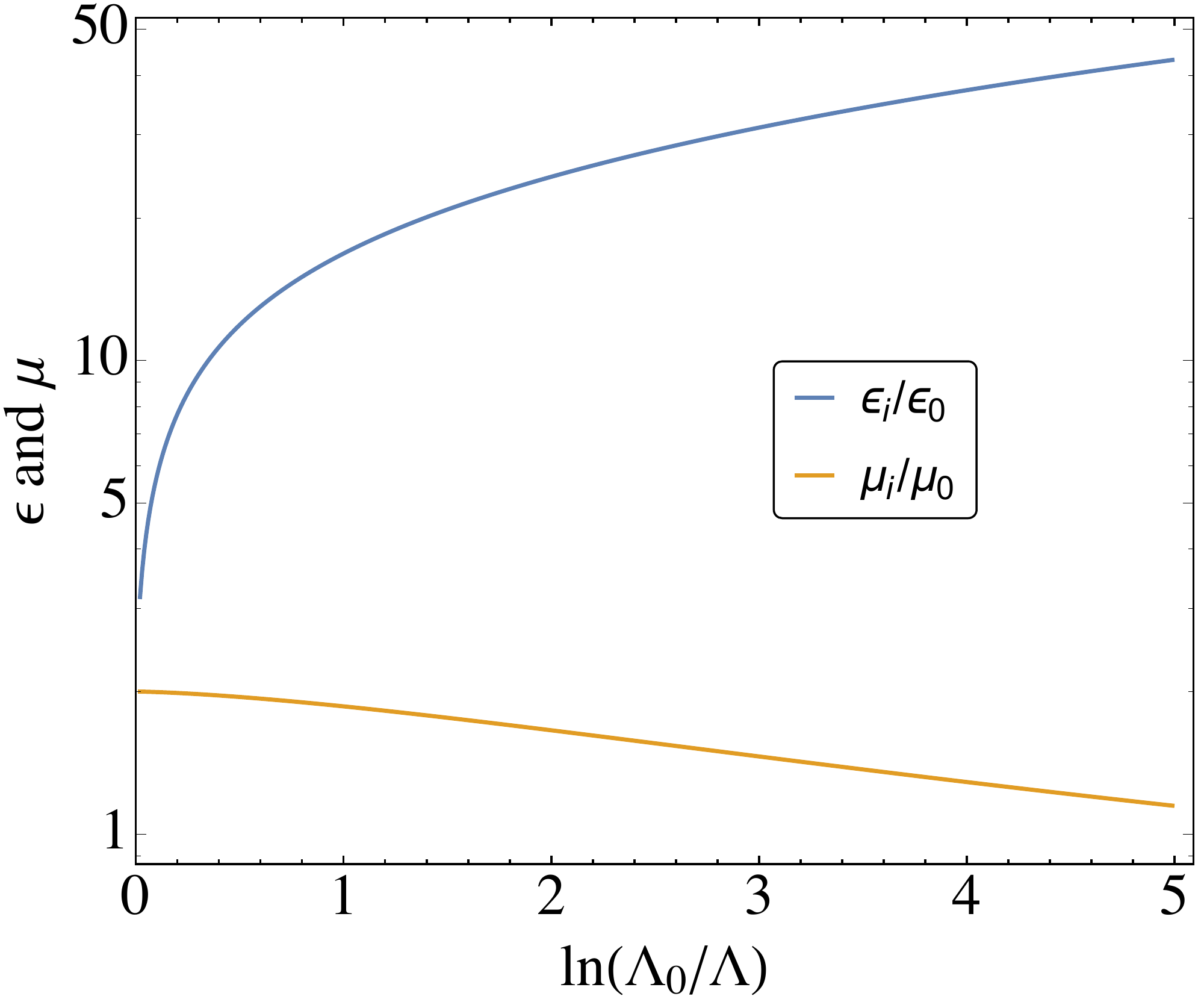}(b)
	\quad
	\includegraphics[width=0.60\columnwidth]{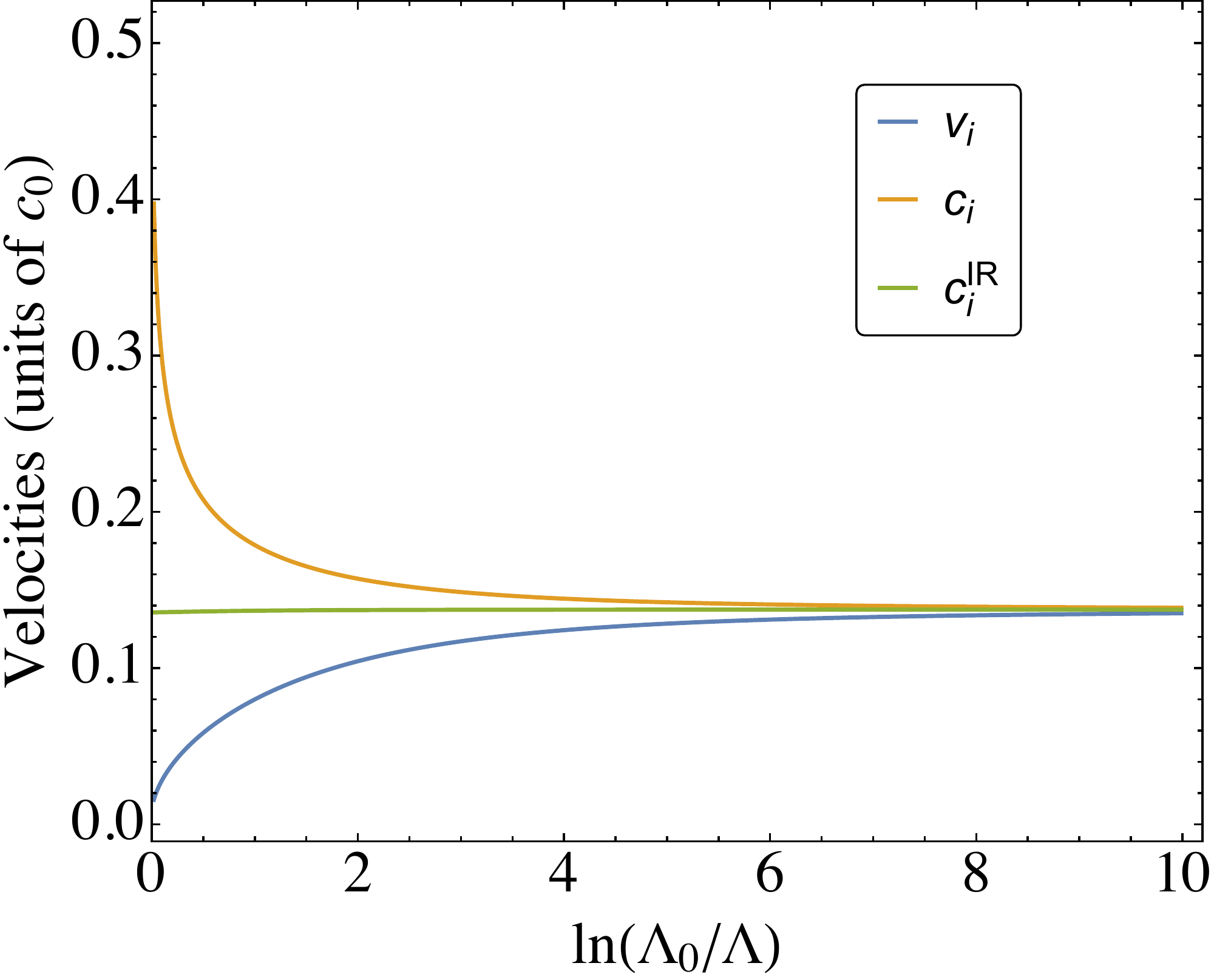}(c)
	\caption{Isotropic runnings of various parameters for the initial values $v_{i}=0.01, \;  \epsilon_{i}=2 , \;  \mu_{i}=2 , \; $ 
with $i=1,2,3$.}
	\label{fig_iso}
\end{figure*}
To fix the notation and as a consistency check we have first analyzed the isotropic case ($v_i=v$) without tilt. This case has already been studied in the literature \cite{IN12,RJH16}. Our results are shown in Fig. \ref{fig_iso} for the initial values given in the caption. The coupling constants go to zero in the infrared limit (Fig. \ref{fig_iso} (a). We took $\alpha=1$ as initial value to compare with the results of ref. \cite{IN12} but the rapid decrease of the couplings ensures the validity of perturbation theory. The velocity of light in the isotropic case is and $c_i=1/\epsilon_{i}\mu_{i}$, $i=1,2,3$.

In Fig. \ref{fig_iso}(b) we show the running of the isotropic electric susceptibility $\epsilon$ and magnetic permittivity $\mu$. In all cases analyzed they run to infinity and zero respectively in the infrared and their product sets the running of the velocity of light $c$. In  Fig. \ref{fig_iso} (c) we show the running of the Fermi velocity and the velocity of light in the material. As we see, they converge to the same non--universal value  which is approximately $c_{i}^{\text{IR}}=(c_{i}^{2}v_{i})^{1/3}$. This agrees with the results in \cite{IN12,RJH16}.

\subsection{Anisotropic Fermi velocity  and no tilt.}
 \begin{figure*}[!]
	\centering
	\includegraphics[width=0.65\columnwidth]{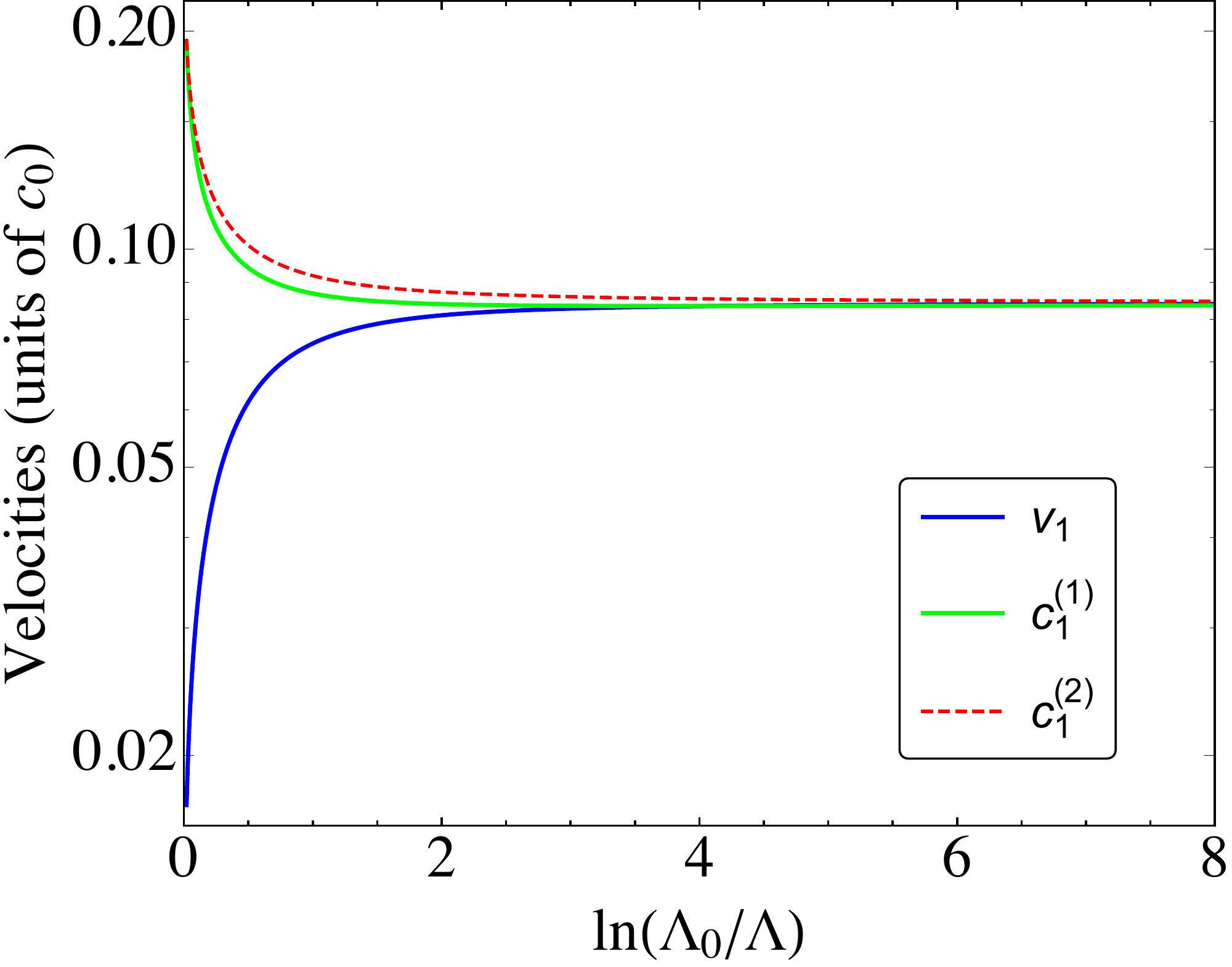}(a)\qquad
	\includegraphics[width=0.65\columnwidth]{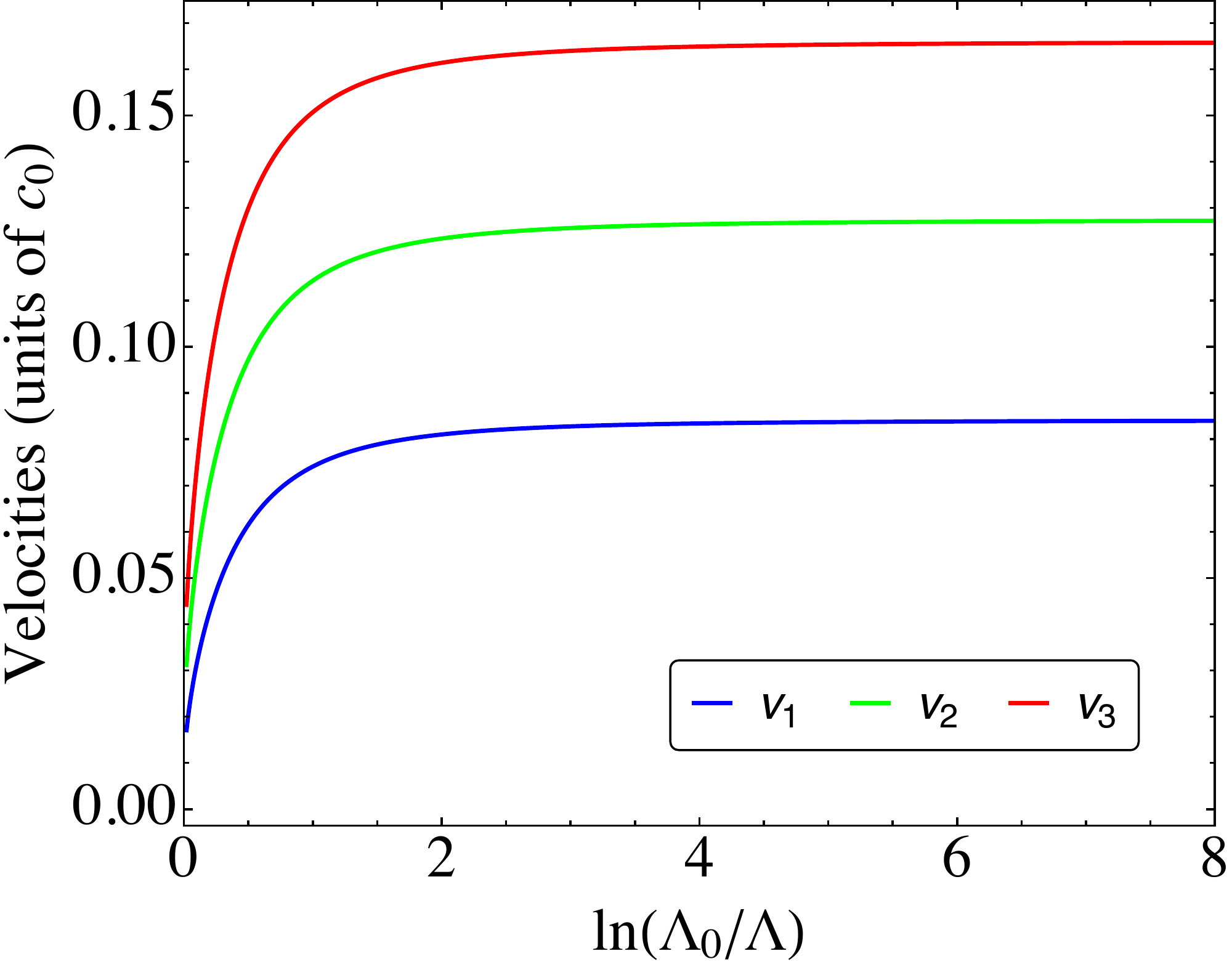}(b)
	\caption{(a) Runnings of the fermion and photon velocities in the X direction  in the anisotropic case with no tilt for the initial values are described in the text. Fermion and photon velocities converge to the same value at the infrared. The same happens in the Y and Z directions. (b) Comparative runnings of the Fermi velocities in the three directions. The asymptotic fixed point is anisotropic.}
	\label{fig_anisNotilt}
\end{figure*}

To our best knowledge this case has not been studied previously in the literature. Surprisingly, rotational symmetry is not recovered at the fixed point. Our results are shown in Fig. \ref{fig_anisNotilt} for the initial values:
\begin{IEEEeqnarray}{rCllrCllrCllrCll}
		\alpha &  \ = \ & 1  ,   \quad &   v_{1} &  \ = \ & 0.01 ,   \quad  & v_{2} &  \ = \ & 0.02    ,   \quad &  v_{3} &  \ = \ & 0.03   ,  \nonumber\\  \epsilon_{i} &  \ = \ & 1   , \quad &  \mu_{1} &  \ = \ & 5    , \quad  &  \mu_{2} &  \ = \ & 10   ,  \quad  &  \mu_{3} &  \ = \ & 15  \    .  \nonumber
\end{IEEEeqnarray}
An analysis of the electromagnetic modes in this  anisotropic case shows that, at the fixed point, there are two transverse modes whose velocities are not given by the simple relation of the isotropic case $c_{i}=c_{0}/\sqrt{\epsilon_{i}\mu_{i}}$ but by the following expressions
\begin{IEEEeqnarray}{rCl}
	c_{1}^{(1)} = \dfrac{c_{0}}{\sqrt{\epsilon_{2}\mu_{3}}}  \  , & \ \ c_{2}^{(1)} = \dfrac{c_{0}}{\sqrt{\epsilon_{3}\mu_{1}}}   \   , &  \ \ c_{3}^{(1)} = \dfrac{c_{0}}{\sqrt{\epsilon_{2}\mu_{1}}}  \  , \\
	c_{1}^{(2)} = \dfrac{c_{0}}{\sqrt{\epsilon_{3}\mu_{2}}}  \  , & \ \ c_{2}^{(2)} = \dfrac{c_{0}}{\sqrt{\epsilon_{1}\mu_{3}}}   \   , & \ \ c_{3}^{(2)} = \dfrac{c_{0}}{\sqrt{\epsilon_{1}\mu_{2}}}  \  . 
	\label{eq_EM}  
\end{IEEEeqnarray}
Their beta functions are given by
\begin{IEEEeqnarray}{rCl}
	\beta_{c_{1}^{(1)}} & = & \dfrac{\alpha_{3}}{3c_{1}^{(1)}}\dfrac{v_{3}}{v_{1}v_{2}}\left( \big( c_{1}^{(1)} \big)^{2}-v_{1}^{2}\left( 1-\dfrac{(t-\omega_{1})^{2}}{v_{3}^{2}} \right) \right)   \   ,   \\
	\beta_{c_{2}^{(1)}} & = & \dfrac{\alpha_{1}}{3c_{2}^{(1)}}\dfrac{v_{1}}{v_{2}v_{3}}\left( \big( c_{2}^{(1)} \big)^{2}-v_{2}^{2} \right)   \   ,   \\
	\beta_{c_{3}^{(1)}} & = & \dfrac{\alpha_{1}}{3c_{3}^{(1)}}\dfrac{v_{1}}{v_{2}v_{3}}\left( \big( c_{3}^{(1)} \big)^{2}-v_{3}^{2}+\left( t-\omega_{1} \right)^{2} \right)   \   ,   
\end{IEEEeqnarray}
for the first propagating mode, and
\begin{IEEEeqnarray}{rCl}
	\beta_{c_{1}^{(2)}} & = & \dfrac{\alpha_{2}}{3c_{1}^{(2)}}\dfrac{v_{2}}{v_{1}v_{3}}\left( \big( c_{1}^{(2)} \big)^{2}-v_{1}^{2} \right)   \   ,   \\
	\beta_{c_{2}^{(2)}} & = & \dfrac{\alpha_{3}}{3c_{2}^{(2)}}\dfrac{v_{3}}{v_{1}v_{2}}\left( \big( c_{2}^{(2)} \big)^{2}-v_{2}^{2}\left( 1-\dfrac{(t-\omega_{2})^{2}}{v_{3}^{2}} \right) \right)   \   ,   \\
	\beta_{c_{3}^{(2)}} & = & \dfrac{\alpha_{2}}{3c_{3}^{(2)}}\dfrac{v_{2}}{v_{1}v_{3}}\left( \big( c_{3}^{(2)} \big)^{2}-v_{3}^{2}+\left( t-\omega_{2} \right)^{2} \right)   \   ,   
\end{IEEEeqnarray}
for the second one. Fig. \ref{fig_anisNotilt}(a) shows that, in the absence of a tilt ($t=\omega_i=0$), $c_{i}^{(1)}=c_{i}^{(2)}=v_{i}$ at the fixed point. Coulomb interaction forces the fermion and photon to propagate identically in the infrared limit at each given direction but the system stays anisotropic. 
In Fig. \ref{fig_anisNotilt}(b) we plot the running of the Fermi velocity components for the given initial values.

\subsection{ Near--isotropic Fermi velocity and non-zero tilt.}
\begin{figure*}[!]
	\centering
	\includegraphics[width=0.65\columnwidth]{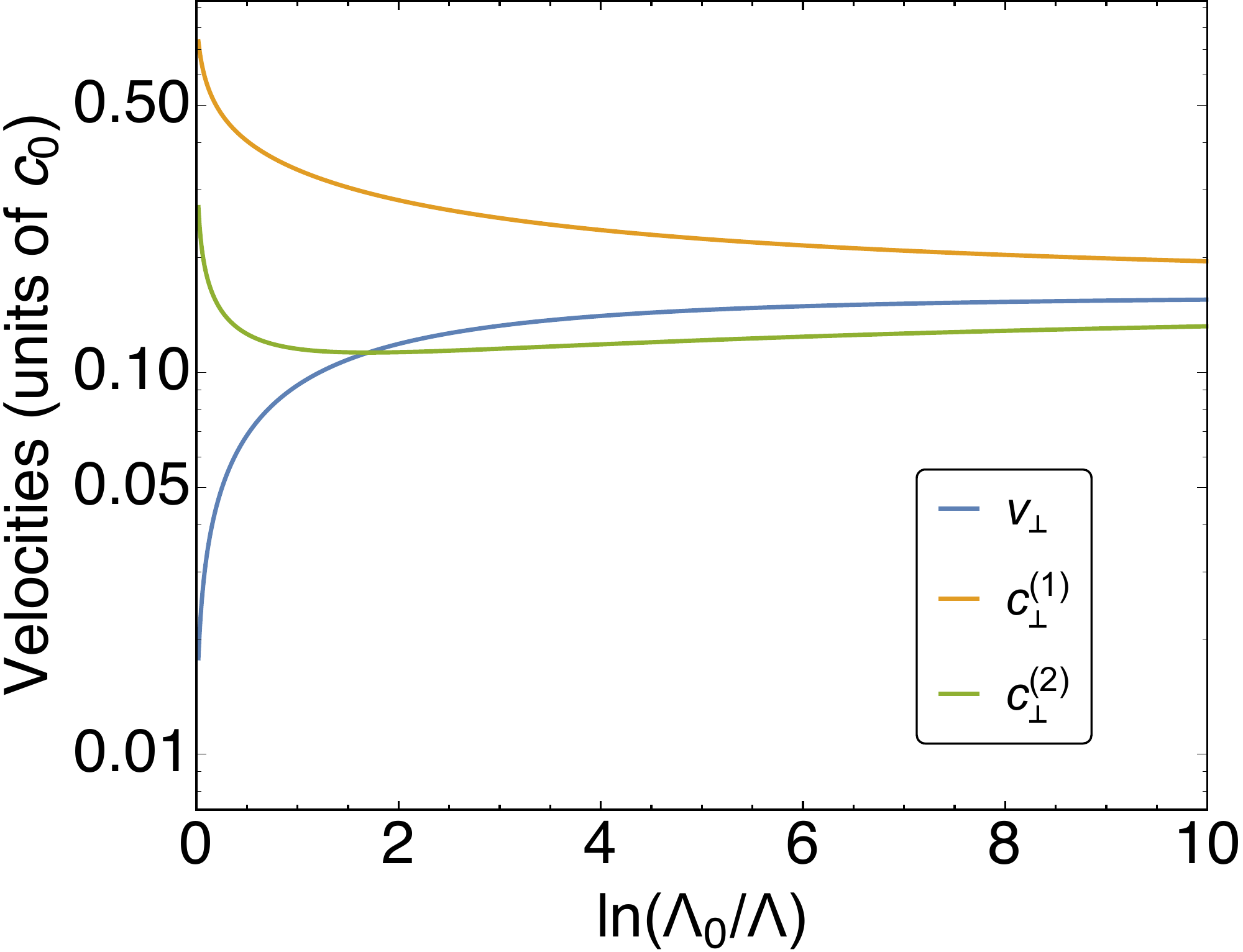}(a)
\qquad
	\includegraphics[width=0.65\columnwidth]{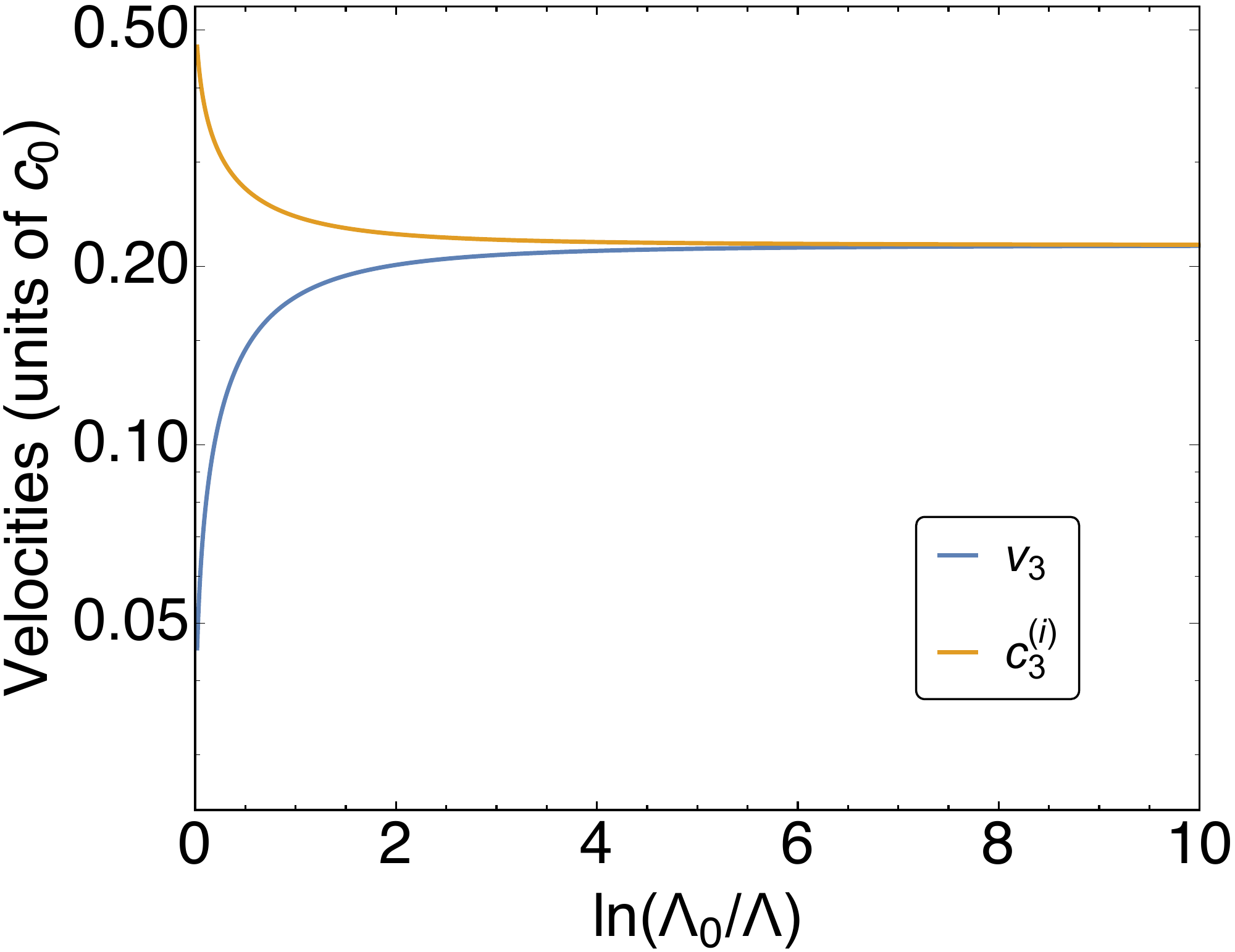}(b)
			\includegraphics[width=0.65\columnwidth]{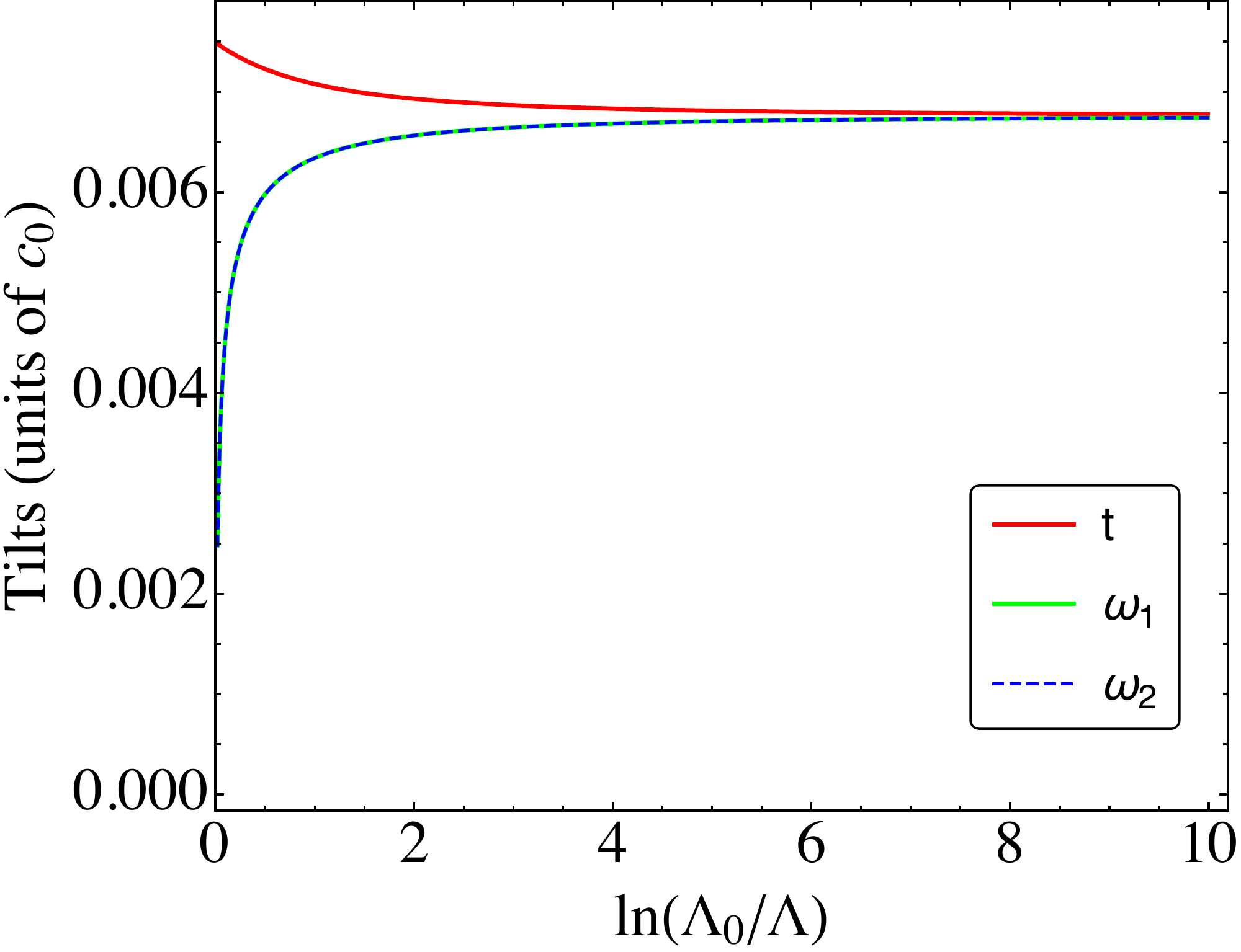}(c)
	\caption{Running of the tilts and velocities for the near-isotropic ($v_1=v_2=0.01$; $v_3=0.03$) case with a fermion tilt $t=t_3=0.0075$. $\epsilon_1=\epsilon_2=1 ; \epsilon_3=3; \mu_1=\mu_2=4; \mu_3=1$.
	(a) and (b) show the running of the fermion and photon velocities in the direction perpendicular and parallel to the tilt. (c) The initial values of the photon tilts $\omega_i$ are set to zero and run to a common value with the fermion tilt at the fixed point. }
	\label{fig_tilt}
\end{figure*}
This is the most interesting case. We begin with a fermion cone tilted in the $Z$ direction and isotropic Fermi velocities in the perpendicular plane.  Fig. \ref{fig_tilt} shows the running of the tilts and velocities for the  initial values given in the caption and $\alpha=1$.
As we see, the  fermion tilt $t_3$ induces photon tilt terms in the perpendicular directions $\omega_{1}$ and $\omega_{2}$, defined in eq. \eqref{eq_tiltPh},  as the energy decreases.  They run to a common value in the infrared fixed point (Fig. \ref{fig_tilt}(c)). The result can be seen analytically, by examining the beta functions 
\beqa
&\beta_{\omega_{1}} = \dfrac{2\alpha_{1}}{3}\dfrac{v_{1}}{v_{2}v_{3}}\left( \omega_{1}-t \right)   ,\quad   \beta_{\omega_{2}}=\dfrac{2\alpha_{2}}{3}\dfrac{v_{2}}{v_{1}v_{3}}\left( \omega_{2}-t \right) \  , \nonumber \\
&\beta_{t}  =  \alpha_{3}\epsilon_{3} \left( tF_{0}^{0}+F_{3}^{0} \right)  \  ,
\eeqa
it is easy to see that there is a fixed point for the $\omega_i$ parameters at $\omega_i=t$. The running of $t$ is less transparent due to the complicated dependence on the parameters of the functions $F_i^j$ defined in 
the Supplemental Material. The physical result is the same irrespective of the initial values of the Fermi velocity components. We have chosen a large initial value for $v_3$ to show an image that magnifies the effect.

An interesting point related to the parameters $\omega_i$ concerns the breakdown of the discrete symmetries ${\cal T}$ and ${\cal I}$ by the tilt velocity of the fermions. 
Since we were interested in the renormalization of the model, we have kept through the work a four dimensional formulation of the massless Lagrangian which describes, generically, a Dirac semimetal with both cones at the same point. Accordingly, the tilt discussed was the same for the two chiralities, what breaks ${\cal I}$. Nevertheless it is possible to introduce an opposite tilt at the two chiralities in which case the model keeps inversion  symmetry. We have checked that, in this case, the contribution to the divergence in the polarization diagrams responsible for the emergence of $\omega_i$ have opposite signs in the two chiralities and cancel, so the magnetoelectric term  is not generated. This was to be expected considering that these terms break ${\cal T}$ and ${\cal I}$. In this particular case, the tilt runs to zero in the infrared. 

\section{Summary and discussion}
\label{sec_discuss} 
The running of the coupling constants in Quantum Field Theory has acquired a richer physical significance with the novel material realization. A particularly important subject for both communities is the fate of the parameters at the fixed point. The restoration of Lorentz invariance has never been questioned before neither in the high energy context nor in the condensed matter. Our result that rotational invariance remains broken at the infrared is  unexpected, and signals an interesting aspect of the differences between  particles propagating  in vacuum and effective low energy models of quasiparticles. Interestingly, this time the difference does not lie on the band structure but on the capability of observing certain quantities as the Fermi velocity.
We have found that the fermion  tilt does not renormalize to zero in the infrared when the polarization function of the full retarded Coulomb interaction is properly taken into account. There are fixed points with finite values of the electronic tilt and, more interestingly, a magneto--electric term that tilts the photon dispersion (eq. \eqref{eq_tiltPh}) is induced when the tilt breaks inversion symmetry.
The various components of the dielectric tensor $\epsilon_{\mu\nu}$  give rise to  birefringence, Cherenkov radiation, Faraday rotation, etc. A very complete analysis of Lorentz-violating modification of electrodynamics was done in ref. \cite{KM09} where magneto--electric terms of the type found in this work have been described. In contrast to what happens in the high energy context, LIVs terms of WSMs are not restricted to very small values and their experimental accessibility selects a preferred frame and allows for the anisotropic fixed point described in this work.
 The propagation of light in WSMs has been addressed in several works \cite{TSetal15,FC16,MC17,FZB17,YKK18}.

\begin{acknowledgments}
We thank K. Landsteiner and A. Cortijo for useful conversations. Y. F. acknowledges support from the ERC Starting Grant No. 679722. This work has been supported by Spanish MECD grant FIS2014-57432-P, the Comunidad de Madrid
MAD2D-CM Program (S2013/MIT-3007), and by the PIC2016FR6. O. P. is supported by an FPU predoctoral contract from MINECO, FPU16/05460.
\end{acknowledgments}
\bibliography{AnisWS}
\appendix
\widetext
\section{Renormalization of the model}
\label{sec_ApRen}
The renormalization functions:
\begin{IEEEeqnarray}{rCl}
	\psi\rightarrow Z_{\psi}\psi  \  , \qquad\qquad & t\rightarrow Z_{t}t  \  , \qquad\qquad & v_{i}\rightarrow Z_{v_{i}}v_{i}  \  ,  \\
	\omega_{i}\rightarrow Z_{\omega_{i}}\omega_{i}  \  ,  \qquad\qquad & \epsilon_{i}\rightarrow Z_{\epsilon_{i}}\epsilon_{i}  \  , \qquad\qquad & \mu_{i}\rightarrow Z_{\mu_{i}}\mu_{i}  \  ,
\end{IEEEeqnarray}
lead to the following diagram contributions
\begin{IEEEeqnarray}{rCl}
	\Sigma_{\text{CT}}(p) & = & i\left( Z_{\psi}-1 \right) \gamma^{0}p_{0}-i\left( Z_{\psi}Z_{t}-1 \right) t\gamma^{0}p_{3}+i\left( Z_{\psi}Z_{v_{1}}-1 \right) v_{1}\gamma^{1}p_{1} \nonumber \\
	& + & i\left( Z_{\psi}Z_{v_{2}}-1 \right) v_{2}\gamma^{2}p_{2}+i\left( Z_{\psi}Z_{v_{3}}-1 \right) v_{3}\gamma^{3}p_{3}  \  ,  \nonumber\\ 
	\Pi_{\text{CT}}^{01}(q) & = & i\left( Z_{\epsilon_{1}}-1 \right) \epsilon_{1}\left( q^{0}q^{1} \right) +i\left( Z_{\epsilon_{1}}Z_{\omega_{1}}-1 \right) \epsilon_{1}\omega_{1}\left( q^{1}q^{3} \right)  \  , \nonumber \\
	\Pi_{\text{CT}}^{02}(q) & = & i\left( Z_{\epsilon_{2}}-1 \right) \epsilon_{2}\left( q^{0}q^{2} \right) +i\left( Z_{\epsilon_{2}}Z_{\omega_{2}}-1 \right) \epsilon_{2}\omega_{2}\left( q^{2}q^{3} \right)  \  ,  \nonumber\\
	\Pi_{\text{CT}}^{03}(q) & = & i\left( Z_{\epsilon_{3}}-1 \right) \epsilon_{3}\left( q^{0}q^{3} \right) -i\left( Z_{\epsilon_{1}}Z_{\omega_{1}}-1 \right) \epsilon_{1}\omega_{1}\left( q^{1}q^{1} \right)  \nonumber \nonumber\\
	& - & i\left( Z_{\epsilon_{2}}Z_{\omega_{2}}-1 \right) \epsilon_{2}\omega_{2}\left( q^{2}q^{2} \right)  \  ,  \nonumber\\
	\Pi_{\text{CT}}^{12}(q) & = & -i\left( Z_{\mu_{3}}-1 \right) \dfrac{1}{\mu_{3}}\left( q^{1}q^{2} \right)  \  ,  \nonumber\\
	\Pi_{\text{CT}}^{13}(q) & = & -i\left( \left( Z_{\mu_{2}}-1 \right) \dfrac{1}{\mu_{2}}+\left( Z_{\epsilon_{1}}Z_{\omega_{1}}^{2}-1 \right) \epsilon_{1}\omega_{1}^{2} \right) \left( q^{1}q^{3} \right)   \   ,  \nonumber \\
	& - & i\left( Z_{\epsilon_{1}}Z_{\omega_{1}}-1 \right) \epsilon_{1}\omega_{1}\left( q^{0}q^{1} \right)  \  ,  \nonumber\\
	\Pi_{\text{CT}}^{23}(q) & = & -i\left( \left( Z_{\mu_{1}}-1 \right) \dfrac{1}{\mu_{1}}+\left( Z_{\epsilon_{2}}Z_{\omega_{2}}^{2}-1 \right) \epsilon_{2}\omega_{2}^{2} \right) \left( q^{2}q^{3} \right)   \   ,   \nonumber\\
& - & i\left( Z_{\epsilon_{2}}Z_{\omega_{2}}-1 \right) \epsilon_{2}\omega_{2}\left( q^{0}q^{2} \right)  \  , 
\end{IEEEeqnarray}
among others. These are the ones we choose to compute the runnings of the parameters, and it has been checked that the other contributions lead to the same set of RG equations. Surprisingly, the vacuum polarization diagram eq. \eqref{VP} has a simple analytic solution for the components we are interested in:
\begin{IEEEeqnarray}{rCl}
	\Pi^{01}(q) & = &  i\dfrac{2\alpha}{3}\dfrac{v_{1}}{v_{2}v_{3}} \left( q^{0}q^{1}+t \ q^{1}q^{3} \right) \dfrac{1}{\varepsilon}  \  ,  \nonumber\\ 
	\Pi^{02}(q) & = &  i\dfrac{2\alpha}{3}\dfrac{v_{2}}{v_{1}v_{3}} \left( q^{0}q^{2}+t \ q^{2}q^{3} \right) \dfrac{1}{\varepsilon}  \  ,  \nonumber\\ 
		\Pi^{03}(q) & = &  i\dfrac{2\alpha}{3}\dfrac{1}{v_{1}v_{2}v_{3}} \left( v_{3}^{2} \ q^{0}q^{3}-tv_{1}^{2} \ q^{1}q^{1}-tv_{2}^{2} \ q^{2}q^{2} \right) \dfrac{1}{\varepsilon}  \  ,  \nonumber\\ 
	\Pi^{12}(q) & = & i\dfrac{2\alpha}{3}\dfrac{v_{1}v_{2}}{v_{3}} q^{1}q^{2} \dfrac{1}{\varepsilon}  \  ,  \nonumber\\ 
	\Pi^{13}(q) & = & i\dfrac{2\alpha}{3}\dfrac{v_{1}}{v_{2}v_{3}} \left( \left( v_{3}^{2}-t^{2} \right) q^{1}q^{3} - t \ q^{0}q^{1} \right) \dfrac{1}{\varepsilon}  \  ,  \nonumber\\ 
	\Pi^{23}(q) & = & i\dfrac{2\alpha}{3}\dfrac{v_{2}}{v_{1}v_{3}} \left( \left( v_{3}^{2}-t^{2} \right) q^{2}q^{3} - t \ q^{0}q^{2} \right) \dfrac{1}{\varepsilon}  \  , 
\end{IEEEeqnarray}
where $\alpha=e^{2}/(4\pi^{2}\hbar\epsilon_{0}c_{0})$ and $1/\varepsilon$ represents the one loop divergence. 

In contrast, the electron self--energy  has to be written in terms of numerical functions. It is convenient to define first $S_{\text{F}}^{\mu\nu}(p-k)\equiv \alpha^{-1}V^{\mu}S_{\text{F}}(p-k)V^{\nu}$ and $S_{\text{F}ab}^{\mu\nu}(k)$ as the term that is proportional to $\gamma^{a}p_{b}$ after Taylor expanding in the external momentum $p$. Then, the term proportional to the pole $\varepsilon^{-1}$ can be computed in terms of an integral of $k_{0}^{E}$ after Wick rotating $k_{0}\rightarrow ik_{0}^{E}$ and another two integrals in spherical coordinates.

Defining the numerical functions
\begin{equation}
   F_{a}^{b}\equiv \dfrac{1}{4\pi^{2}}\int_{-\infty}^{\infty}dk_{0}^{E}\int_{0}^{\pi}d\theta\sin\theta \int_{0}^{2\pi}d\varphi \ k^{3}S_{\text{F}ab}^{\mu\nu}(k_{0}^{E},k,\theta,\varphi)G_{\mu\nu}(k_{0}^{E},k,\theta,\varphi)  \  ,
\label{eq_F}
\end{equation}
the electron self--energy is given by
\begin{IEEEeqnarray}{rCl}
	\Sigma(p) & \equiv & \Big( i\alpha F_{0}^{0}\gamma^{0}p_{0}+i\alpha F_{3}^{0}\gamma^{0}p_{3}+i\alpha F_{1}^{1}\gamma^{1}p_{1} \nonumber \\ 
	& +& i\alpha F_{2}^{2}\gamma^{2}p_{2}+i\alpha F_{3}^{3}\gamma^{3}p_{3}\Big) \dfrac{1}{\varepsilon}  \  .
\label{eq_Fs}
\end{IEEEeqnarray}
Although we have chosen not to renormalize the electric charge, for completeness we quote the one--loop vertex funtion which is given by: $\Gamma^{0}(0,0)  \equiv  -i\alpha F_{0}^{0} e \gamma^{0}$.

\end{document}